\documentclass[proceedings]{JHEP3}

\PrHEP{PrHEP hep2001}                   
\conference{International Europhysics Conference on HEP}                

\usepackage{epsfig}                   

\title{What we can learn on inflation from recent
\linebreak CMBR data}

\author{\speaker{Laura Covi} 
        \\  
        DESY, Deutsches Elektronen Synchrotron, Theory Group, 
        Notkestrasse 85, D-22603~Hamburg, Germany \\    
        E-mail: \email{Laura.Covi@desy.de}}                       

\abstract{We review the prediction of inflation and 
the constraints on inflationary models
coming from recent observations.}

\begin{document}

\section{Introduction}

Inflation is a period of exponential expansion of the scale
factor of our universe, supposed to have taken place before the standard 
hot Big Bang cosmology. It is necessary to explain the homogeneity, 
isotropy and flatness of the present universe and also the absence 
of unwanted relics, some of the questions not solved by the standard 
picture \cite{infl-book}.

\subsection{Generic predictions}

The inflationary paradigm, independently of the specific model, 
makes very powerful predictions \cite{infl-book}:

\begin{itemize}

\item{the universe is flat with very high precision, i.e. the
total energy density is equal to the critical one, $\Omega_{tot} =1$; 
in fact during slow roll inflation, when the scale factor
$a(t) \propto e^{H t} $, with practically constant Hubble parameter 
$H= \dot a/a$, we have
\begin{equation}
{1\over \Omega_{tot} (t_{end})} = 
1 - \left(1-{1\over \Omega_{tot} (t_{in})}\right) e^{-2 N} \, ,
\end{equation}
so that the total energy density tends exponentially 
towards the critical density for large e-folding number 
\begin{equation}
N = \int_{t_{in}}^{t_{end}} H dt \,\geq\, 60\, ;
\end{equation}
}
\item{in the simple single field case,
the primordial perturbations are gaussian and adiabatic;
the gaussianity is related to the fact that the perturbations
are originated by the quantum fluctuations of the 
inflaton, and it is the reason why all the 
information on the perturbations is encoded in their power 
spectrum;}
\item{the spectrum of the perturbations is nearly scale invariant
due to the slow rolling of the inflaton field, and the deviation
from scale invariance are a characteristic of the model, as we will 
see.}

\end{itemize}

\subsection{Model--dependent predictions}

In the simplest implementation, a model of inflation consists in a 
scalar potential $V(\phi)$ for the inflaton field $\phi $ satisfying 
slow roll conditions 
\cite{lr99}:
\begin{eqnarray}
\varepsilon & = & {1\over 2} M_P^2\left({V'\over V}\right)^2 \ll 1\\
|\eta |        & = & M_P^2\left|{V''\over V}\right| \ll 1
\label{slowroll}
\end{eqnarray}
where $M_P=2.4\times 10^{18}\,\mbox{GeV}$ is the reduced Planck mass and 
the prime denotes derivative with respect to the field $\phi$.

The quantum fluctuations of the inflaton field on the classical
background generate a primordial gaussian perturbation of the 
curvature tensor, 
which can be the origin of the large scale structure in the Universe. 
The point of contact between observation and models of inflation
is the Fourier transform of the perturbation in comoving momentum space, 
or more precisely its power spectrum ${\cal P}_{\cal R}(k)$, which, 
in the slow roll approximation $3H\dot\phi\simeq -V'$, 
 is given in terms of the inflaton potential $V(\phi)$ by
\begin{equation}
{\cal P}_{\cal R}(k)
 = \left. \frac{1}{12\pi^2 M_P^6}\frac{V^3}{V'^2} \right|_{k=aH} \,,
\label{delh}
\end{equation}
where the potential and its derivatives are evaluated at the epoch of 
horizon exit $k=aH$. To work out the value of $\phi$ at this epoch 
one uses the relation
\begin{equation}
\ln\left({k_{\mathrm end}\over k}\right)\equiv N(k)
=M_P^{-2}\int^\phi_{\phi_{\mathrm end}} {V\over V'} d \phi
\,,
\label{Nofv}
\end{equation}
 where $N(k)$ is the actual number of $e$-folds from horizon exit of
the scale $k$ to the end of slow-roll inflation.
The e-folding number at the scale explored by the COBE DMR 
experiment~\cite{cobe-exp} measuring the cosmic microwave background 
radiation~(CMBR) anisotropy, 
 $N(k_{cobe})$, depends on the expansion of the Universe after inflation
in the manner specified by:
\begin{equation}
N(k_{cobe}) \simeq 60 - \ln(10^{16}\, \mbox{GeV}/V^{1/4}) - \frac{1}{3}
\ln(V^{1/4}/T_{reh}) 
\, .
\label{Ncobe}
\end{equation}
In this expression, $T_{reh}$ is the  reheat temperature, and
instant reheating is assumed. 

Given the above relations, the observed large-scale normalization  
measured by COBE DMR \cite{cobe}
\begin{equation}
{2\over 5} {\cal P}_{\cal R}^{1/2} = \delta_H (k_{cobe}) = 2.1\times 10^{-5}
\label{cobenorm}
\end{equation} 
provides a strong constraint on models of inflation.

Another important information is contained in the scale-dependence of
the spectrum, defined by the, in general, scale-dependent spectral 
index $n$;
\begin{equation}
n(k)-1\equiv \frac{d \ln {\cal P}_{\cal R}(k)}{d\ln k}\,.
\end{equation}

From observations, we know that $n$ is very near to one.
According to most inflationary models, $n$ has very small variation on
cosmological scales, since
\begin{equation}
\frac{d n(k)}{d\ln k} \propto (n(k)-1)^2 \ll 1\, ;
\end{equation}
then we can write the power-law formula 
${\cal P}_{\cal R} (k)\propto k^{n-1}$, which reduces to the
scale-invariant Harrison-Zeldovich form for $n=1$.
But in general the dependence on the scale can be much stronger.

From (\ref{delh}) and (\ref{Nofv}),
\begin{eqnarray}
n-1 &=&  2 M_P^2 (V''/V)-3 M_P^2 (V'/V)^2 \,,
\label{nofv}
\end{eqnarray}
and in all models where inflation takes place near a (local) maximum
or minimum of the potential, (\ref{nofv}) 
is well approximated by
\begin{equation}
n-1=2 M_P^2(V''/V)\,.
\label{nofvapprox}
\end{equation}
We see that the  spectral index 
 measures the {\em shape} of the inflaton potential $V(\phi)$,
being independent of its overall normalization. For this reason, 
it is a powerful discriminator between models of inflation.

Analogously to the scalar perturbations, also tensor perturbations
are generated by the quantum oscillations of the inflaton field.
For those, the power spectrum is given by
\begin{equation}
{\cal P}_{grav} (k) =  \left. {V \over 6 \pi^2 M_P^4} \,\right|_{k=aH}
\end{equation}
and the spectral index is
\begin{equation}
n_{grav}(k) =  \left. {d\log({\cal P}_{grav})\over d\log(k)}  
\right|_{k=aH} = - 2 \epsilon\, .                 
\end{equation}

Note that the power spectrum of tensor perturbations is 
much smaller than the scalar one, since we have
\begin{equation}
{{\cal P}_{grav} (k) \over {\cal P_R} (k)} 
=  \epsilon \ll 1\, ;
\end{equation}
this gives for the CMBR anisotropy the tensor to scalar ratio 
at low $\ell$ \cite{r-quad}
\begin{equation}
r \equiv {C_\ell^{grav} \over C_\ell^{\cal R}} 
\simeq 12.4 \, \epsilon\, .
\end{equation}
The tensorial contribution to the CMBR anisotropy, present at 
large scales, is for this reason subdominant or even
completely negligible for models with very small $\epsilon $, as those
we will consider.

Tensor perturbations can be detected independently from the scalar
one through the polarization of the CMBR \cite{pol} or 
through gravitational waves (but their level are unfortunately 
well below the present and future experimental sensitivities of
gravitational waves detectors \cite{gravdet}).

We will describe in the next section as examples a couple of models
of inflation and then review the present constraints.


\section{Models of inflation, some examples}

Let us now describe in particular hybrid inflation \cite{Linde-h}. 
It is a two-field model, where one of the fields is the inflaton
and the second, the hybrid field, is responsible of a phase
transition at the end of inflation, but is static during inflation.
\EPSFIGURE{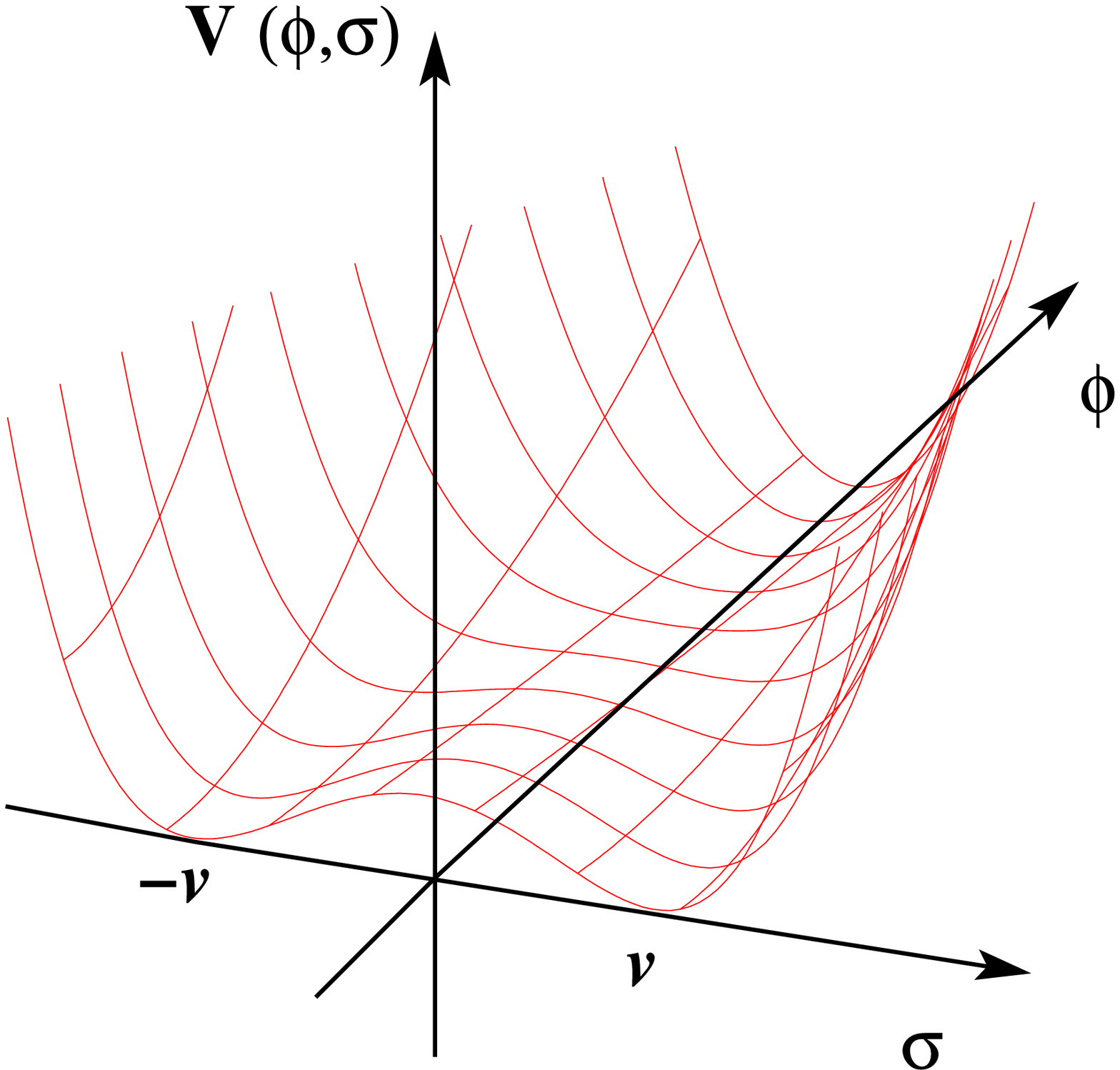,width=7cm,height=7cm}%
{\it Potential for hybrid inflation.\label{fig1}}   
The scalar potential for this kind of model looks like
\begin{equation}
V(\phi,\sigma) = \kappa (\sigma^2 - v^2)^2 + \kappa' \phi^2 \sigma^2 +
V_\phi (\phi)\, ;
\end{equation}
during inflation $\phi^2 \geq {2\kappa\over \kappa'} v^2 $ and the hybrid
field is stabilized at the origin, so that the potential driving inflation, is
\begin{equation}
V(\phi,0) = V_0 + V_\phi (\phi)\, ,
\end{equation}
with $V_0 = \kappa v^4 $; this expression has to be considered to compute the 
power spectrum
and spectral index, as described in the previous section.
Different hybrid inflationary models arise depending on the choice of 
$V_\phi (\phi)$.
The typical shape of the hybrid inflationary potential is shown in Fig. 1,
while some prediction for the spectral index on different models are
shown in Table 1.

\TABLE{
\begin{tabular}{@{}lclp{5cm}}
\hline\noalign{\smallskip}
$1+ V_\phi(\phi)/V_0$ & $n-1$ & $d n/d\log(k)$ & Origin of the slope \\
\noalign{\smallskip}
\hline
\noalign{\smallskip}
$1-\frac{\lambda^2}{4\pi^2} \log\left({\sqrt{2}\lambda\phi\over Q}\right)$ 
& ${-1\over N+\frac{2\pi^2\phi_c^2}{\lambda^2 M_P^2}}$ &
$ {-1\over\left(N+\frac{2\pi^2\phi_c^2}{\lambda^2 M_P^2}\right)^2}$ &
1 loop for spont. broken susy\\
\noalign{\smallskip}
$ 1\pm \frac{1}{2} \frac{m^2_\phi}{V_0} \phi^2$ & 
$\pm \frac{m^2_\phi M_P^2}{V_0}$ & $0$
& Susy breaking mass \\
\noalign{\smallskip}
$1 \pm \beta \frac{\phi}{M_P} $& $- 3\beta^2$ & $0$ &
Susy breaking linear term\\
\noalign{\smallskip}
$1 \pm \frac{\phi^4}{M_P^4} $ & ${12\over \frac{M_P^2}{2 \phi_c^2}\mp N}$&
  ${\mp 12\over \left(\frac{M_P^2}{2 \phi_c^2}\mp N\right)^2}$ &
Sugra quartic term\\
\noalign{\smallskip}
\hline
\noalign{\smallskip}
\end{tabular}
\caption{Various models of hybrid inflation and their prediction
for the spectral index as a function of $N= -\log(k/k_{\mathrm end})$
and its derivative, assuming small field values. 
In the case of the simple linear term, $n-1$ is given by the full
expression (\ref{nofv}) since $\eta $ vanishes, but note that
in supergravity generally, other contribution to the spectral index
coming from higher order terms are usually present and can be larger
than the one listed here \cite{bcd00}.}
\label{LC-Tab1}
}

We will describe in the following a couple of examples of hybrid
inflation constructed within local supersymmetric theories.
While supersymmetry is a vital ingredient at low energy for solving
the hierarchy problem and stabilizing scalar masses, it could seem
unnecessary to invoke it during inflation, especially since 
such symmetry is explicitely broken by the large effective cosmological 
constant responsible of the inflationary phase.
It turns out anyway that supersymmetry brings many advantages also
to inflationary model building, not only stabilizing the inflaton
potential and its small parameters, but also providing many scalars
as inflaton candidates, in particular the flat directions of the scalar
potential. 
Moreover, if we assume low energy supersymmetry, as required
by the hierarchy problem, it should certainly not be neglected
at the large scale when inflation takes place.

However, we must not forget the fact that the
inflationary vacuum energy breaks strongly supersymmetry and
that supergravity corrections can play an important role \cite{lr99}.

\subsection{Linear term hybrid inflation in supergravity}

We will consider a model of inflation with superpotential \cite{bcd00}
\begin{equation}
W = \lambda T \left(M_G^2 - \Sigma^2\right) + M_S^2 (\beta + S)\;,
\label{W-infl}
\end{equation}
where $T, S$ and $\Sigma$ are chiral superfields. The second part
of the superpotential is the Polonyi potential \cite{po77} and
allows for supersymmetry breaking in the true vacuum with
vanishing cosmological constant for  $\beta \simeq (2-\sqrt{3}) M_P$.
$M_S$ is then the supersymmetry breaking scale, yielding the gravitino mass 
$m_{3/2} \simeq M_S^2/M_P $.

For global supersymmetry the scalar potential reads
\begin{equation}
V  = \lambda^2 |M_G^2-\Sigma^2|^2 + 4\lambda^2 |T|^2 |\Sigma|^2 + M_S^4\;,
\label{linear-V}
\end{equation}
and it is of the hybrid inflationary type, even if perfectly flat along
the $T$ and $S$ directions.
A small curvature needed for the `slow roll' along the $T$ direction
is generated by many contributions, e.g. quantum corrections due to the 
loops of the $\Sigma$ particles \cite{dss94}, and also supergravity
corrections. 

For large $T$ and zero $\Sigma $ the dominant corrections to the potential
give \cite{bcd00}
\begin{equation}
V(\phi ) =
\lambda^2 M_G^4 \left[ 1 - 2\sqrt{2}\xi\,{\beta \,\phi \over M_P^2}
+ {\phi^4\over 4 M_P^4} +
{\lambda^2 \over 8\pi^2} \log \left({2\lambda^2 \phi^2\over \mu^2}\right)
+ ... \right].     
\end{equation}
where $\xi = M^2_S/ (\lambda M^2_G) \ll 1$, the first two terms come
from supergravity corrections, and the last from one loop radiative
corrections.
The three terms in the potential compete and depending on the parameters 
$\lambda $, $\xi$ and $M_G$, different regimes can be realized. 
In all cases we
have a viable model of hybrid inflation and we can fix one of the
three parameters using the COBE normalization (\ref{cobenorm}).
Taking the supersymmetric scale $M_S$ to be in a phenomenologically
acceptable range for low energy supersymmetry, 
$M_S = 1.4\times 10^{10} \mbox{GeV} $ the parameter space is shown in Fig. 2.
\EPSFIGURE{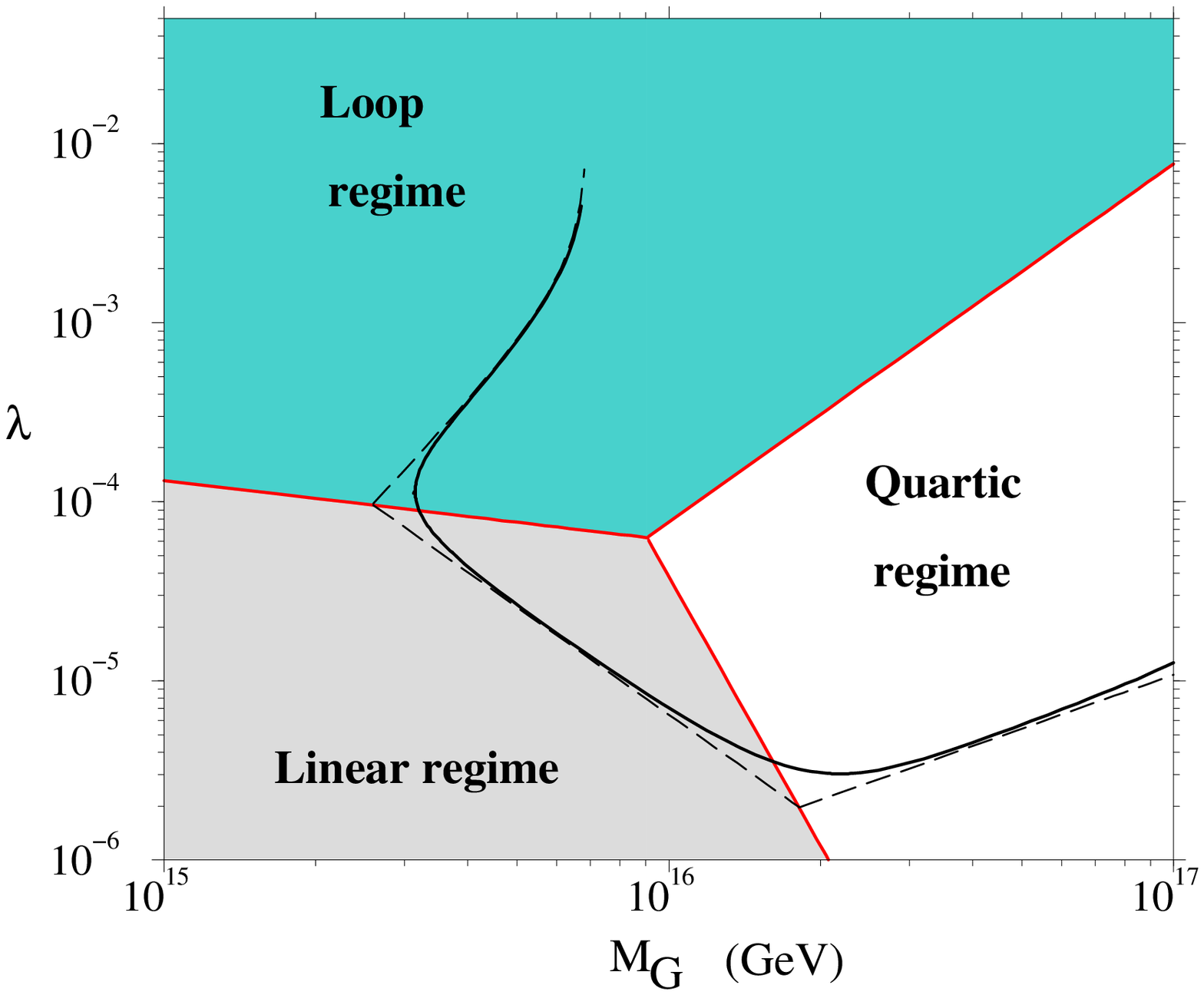,width=8cm}%
{\it The three regimes of hybrid inflation: the loop regime, the
linear regime and the quartic regime. The COBE normalization defines a curve
(full line) in the $\lambda - M_G$ - plane for 
$M_S = 1.4\times 10^{10}\, {\mathrm GeV}$. The dashed lines are obtained by
keeping only the dominant term in the inflaton potential.
}

We see that the linear term dominates for small couplings
and small $M_G$, and in that regime, from the COBE normalization 
(\ref{cobenorm}), one obtains, 
\begin{equation}
{1\over 2\sqrt{150}\pi}{M_S^2 \over \xi^2 \beta M_P} 
= 2.1 \cdot 10^{-5}\;.
\end{equation}
For $M_S \simeq  10^{10}\mbox{GeV}$, this gives \linebreak
$\xi \equiv M^2_S/(\lambda M_G^2) \simeq 10^{-6}$. Note, that 
$\xi$ is the ratio of the gravitino masses in the true vacuum and in the
inflationary phase. Since $\xi \ll 1$, slow-roll conditions are well
satisfied for $\phi \simeq \phi_c$, 
and the spectral index is
\begin{equation}
n -1 \simeq 3 {\phi_*^2\over M_P^2} \leq  2.4 \cdot 10^{-4}\;.\label{n-1}
\end{equation}
An inflationary phase dominated by a linear term is very interesting, since 
it gives a {\it scale invariant spectrum} to high accuracy. For standard
hybrid inflation, on the contrary, one has $n\simeq 0.98$ \cite{dss94}. 
Future satellite 
experiments may eventually be able to distinguish between these two regimes
of hybrid inflation.

\subsection{Mass term hybrid inflation}

In case the inflaton transforms under a discrete or continuous
symmetry, the linear term in the potential vanishes due to the
symmetry and the lowest order term in the scalar potential lifting 
a flat direction is a mass term. We have in that case
\begin{equation}
V_\phi (\phi) = {1\over 2} m^2 \phi^2 + ...\, ;
\end{equation}
the dots are higher order terms who can become important at large
field values.

Many models of this type, based on supersymmetric extensions of the
Standard Model have been considered, starting with \cite{rsg95}.
These models can have many different feature, depending on the
particular implementation, see \cite{lr99} for a review.
For example, it has been recently realized, that in this class of
models inflation can take place even at very low scales \cite{grs01}, 
softening the problem of realizing an inflationary epoch in the case 
of large extra-dimensions when the Planck mass is of order of the
TeV scale.

Another interesting signature is the peculiar strong
scale dependence of the spectral index, that appears in the case
of large one loop quantum corrections to the inflaton potential
\cite{st97}. Resumming those terms into the inflaton mass, we have
in the potential the running mass $m(\phi) $: 
\begin{equation}
V_\phi (\phi) = {1\over 2} m^2(\phi) \phi^2 + ...\, ;
\end{equation}
so that, assuming the end of inflation is set by the beginning of fast roll,
 the spectral index is given by \cite{lc00}
\begin{equation}
{n(k)-1\over 2} \simeq  \left. m^2(\phi)\right|_{k=aH} \,
=  s \left({k\over k_{cobe}}\right)^c - c  \, ;
\label{run-n}
\end{equation}
where $s$ is an integration constant related to $\phi_{end}$,
$c \propto {d m^2 \over d\log(Q)} $ is proportional to the 
beta-function of the inflaton mass. 
We see that the scale dependence of the spectral index is in this
case pretty strong (a power-law, but for the spectral index itself 
not the power spectrum~!).

\section{Comparison to observations}

To obtain information about the primordial power spectrum
and so on inflation, it is necessary to follow the 
perturbations from the inflationary to the present epoch 
and compare the processed spectrum to the observed one. 
Unfortunately, the evolution of the perturbations 
depends on the background cosmology and therefore on the 
cosmological parameters, i.e. the present Hubble parameter
in units of $100\, \mathrm{ km/s/Mpc}, h$, 
the total energy density, $\Omega_{tot}$,
the matter density $\Omega_M $ and the nature and composition
of Dark Matter, the baryon density $ \Omega_b$.
It is for this reason not so straightforward to gain 
information on the primordial spectrum, without any
assumption on the cosmology, or as it is usual said,
without {\it biases}.

On large scales, below the size of the present horizon 
$ \approx 3000\, \mathrm{Mpc} $ down to around 
$10\,\mathrm{Mpc}$,  information on the power spectrum
is provided by the CMBR anisotropies, on smaller scales, 
about $100$--$1\,\mathrm{Mpc} $ information comes 
instead from the visible matter power spectrum.
During radiation dominance, due to the presence of pressure,
the perturbations could not grow nor reach the instability regime
and the dynamics was just an oscillation in the radiation plasma;
the CMBR anisotropies are a snapshot of this period and they
provide us with the cleanest signal. In fact, the dynamics of the
plasma at that epoch is well understood and nowadays powerful
computer codes like CMBFAST \cite{cmbfast}  are available to the 
scientific community to compute the CMBR anisotropies specifying 
the initial conditions and compare them to observations.
On the other hand extracting the primordial power spectrum
from the present matter power spectrum is more complicated
since such perturbation reached the non-linear regime and 
underwent gravitational collapse.

\subsection{CMBR observations}

This year three new experimental measurements of the CMBR
anisotropy were completed, providing the first view of 
three acoustic peaks in the CMBR spectrum \cite{Max01, Boom01, Dasi01}.

The analysis of these data in order to extract the cosmological
parameters have been performed, both for the single data sets,
\cite{Dasi01bis, Max01bis, Boom01bis}, and for the combined data 
\cite{wtz01}, including also Maxima, Boomerang, Dasi and CBI \cite{CBI}.
All the analysis give results that are in good agreement with
the general prediction of inflation:
\begin{itemize}
\item{the position of the first peak of the CMBR is a direct measurement
of the geometry of our universe and give a very clear indication
that the spatial curvature is vanishing in accordance with the
inflationary prediction; the total energy density as inferred
from the CMBR only \cite{wtz01} is:
\begin{equation}
\Omega_{tot} =  1.06^{+0.13}_{-0.59} \, .
\end{equation}

As we can see, the precision of the measurement is not very high,
due to degeneracies between the cosmological parameters, but
imposing even weak biases on the values of some parameters or
considering also large scale structure data, reduces 
strongly the uncertainties: 
}
\TABLE{\begin{tabular}{@{}lll}
\hline\noalign{\smallskip}
$\Omega_{tot} $ & Bias & Data and reference\\ 
\noalign{\smallskip}
\hline
\noalign{\smallskip}
 $1.04^{+0.11}_{-0.12}$ &  $ h> 0.45,\;\tau_c < 0.4$ 
& DMR \& DASI \cite{Dasi01bis}\\
 $0.90^{+0.18}_{-0.16}$ & $0.4<h<0.9,\;\Omega_M > 0.1,\; t_0 > 10$ Gyr
& DMR \& MAXIMA-I \cite{Max01bis}\\
 $1.02^{+0.10}_{-0.10}$ & $0.45<h<0.85,\; t_0 > 10$ Gyr
& DMR \& BOOMERANG \cite{Boom01bis}\\
 $1.00^{+0.06}_{-0.06} $& $ h=0.72\pm 0.08 $  
& PSCz \& combined CMBR data \cite{wtz01} 
\end{tabular}%
\caption{Results for the total energy density of the
universe with different biases and datasets.
The errors correspond to 95\% CL. For the third line \cite{Boom01bis}, 
the error is obtained reading Fig.~4.\label{Omegatot}}}

\item{it seems also that the curvature perturbations are
compatible with a primordial spectrum gaussian \cite{gauss}
and adiabatic \cite{adiab}, so that we can conclude that
the single field slow roll inflation does give a good 
fit to the present observations;}
\item{a nearly scale invariant primordial spectrum
${\cal P}_{\cal R} (k)\propto k^{n-1}$ with $n\simeq 1$ is a
good fit of the data, as we will see below in more detail;}
\item{note also that the new CMBR observations are in very good 
agreement with other independent measurements, e.g. with the 
value of the baryon density obtained from Nucleosynthesis, 
$\Omega_b h^2 = 0.020 \pm 0.002$ \cite{BBN00}, and the
determinations of the matter density from astrophysical 
observations \cite{omegam}. The data seem also to prefer
very low values for the hot dark matter density \cite{wtz01}.}
\end{itemize}

On the other hand, present data are much less powerful
in discriminating between different models. For the spectral
index we have in fact still a pretty wide allowed interval,
containing most slow-rolling models, even after imposing
constraints on the cosmological parameters:
\TABLE{\begin{tabular}{@{}lll}
\hline\noalign{\smallskip}
$ n $ & Bias & Data and reference\\ 
\noalign{\smallskip}
\hline
\noalign{\smallskip}
 $1.01^{+0.16}_{-0.11}$ &  $ h> 0.45,\;\tau_c < 0.4$ 
& DMR \& DASI \cite{Dasi01bis}\\
 $0.99^{+0.14}_{-0.14}$ & $0.4<h<0.9,\;\Omega_M > 0.1\; ,t_0 > 10$ Gyr, 
$\,\tau_c = 0$ & DMR \& MAXIMA-I \cite{Max01bis}\\
 $0.97^{+0.18}_{-0.18} $ & $0.45<h<0.85,\; t_0 > 10$ Gyr
& DMR \& BOOMERANG \cite{Boom01bis}\\
 $0.93^{+0.13}_{-0.10} $& $ h=0.72\pm 0.08 $  
& PSCz \& CMBR data \cite{wtz01} 
\end{tabular}%
\caption{Results for the spectral index using different biases and datasets.
The errors on $n$ correspond to 95\% CL, and again for \cite{Boom01bis}, 
the 95\% CL interval is obtained from Fig. 4.\label{ennefit}}}

One of the problems of extracting $n$ is due to the degeneracy with
the optical depth to the surface of last scattering $\tau_c$; such
degeneracy can be reduced modeling reionization and
estimating $\tau_c$ from the power spectrum itself \cite{lc00}.

In all these analysis, the primordial power spectrum is assumed to
be a power-law with a constant spectral index. This is a reasonable
assumption for many inflationary models, but it leaves unanswered
an important question, i.e. how large is the scale dependence in 
$n$ allowed by the data. 
Two groups have studied this kind of constraints, but unfortunately 
have not yet up-dated their analysis to consider the latest data.
In \cite{lc00} the specific scale dependence of running mass models
given by eq. (\ref{run-n}) has been compared to observations and
constraints on the values of the $s$ and $c$ parameters have been 
obtained, e.g. $-0.31 \leq s \leq 0.21 $ and 
$-0.21 \leq c \leq 0.15$ at 95\% CL.
The analysis in \cite{hann00,hann01} relies instead in a Taylor expansion
of the power spectrum as a function of $\log (k)$, truncated after
the first derivative of the spectral index. Using not only CMBR and
LSS data, but also linear matter power spectrum from Ly-$\alpha$
forest spectra, \cite{hann01} obtained the strong constraint
$ -0.05 \leq {dn\over d\log(k)} \leq 0.02$ at the 2$\sigma$ level.

\section{Conclusions}

We have seen that the single field inflationary paradigm is very 
successful in describing present observations, but unfortunately
the precision of the present data is not yet sufficient to
discriminate between the explicit models. To extract information 
on the primordial power spectrum from the CMBR is necessary to
exploit all our knowledge of the cosmological parameters in order
to reduce the degeneracies.

It is foreseeable that in the future a much better determination
of the spectral index $n$ will be achieved, thanks both to more
precise satellite experiments like MAP \cite{map} and to the 
improvement of the measurements of the cosmological parameters
by other astrophysical methods.



\section*{Acknowledgments}

The author is grateful and indebted to her collaborators, 
W.~Buchm\"uller, D.~Del\'epine and D.~H.~Lyth.
She would like to thank the Conference organizers for local 
financial support and the parallel session conveners for the 
opportunity to speak.


\begin{thebibliography}{99}

\bibitem{infl-book} 
A.~D.~Linde, {\it Particle Physics and Inflationary Cosmology},
Harwood Academic (1990)

\bibitem{lr99}
For a review and references, see
D.~H.~Lyth and A.~Riotto, \prep{314}{1999}{1} 

\bibitem{cobe-exp}
C.~L.~Bennett et al., \apj{464}{1996}{L1}

\bibitem{cobe}
E.~F.~Bunn and M.~White, \apj{480}{1997}{6}

\bibitem{r-quad} 
A.~R.~Liddle and D.~H.~Lyth, \prep{231}{1993}{1}

\bibitem{pol} M.~Kamionkowski and A.~H.~Jaffe, astro-ph/0011329

\bibitem{gravdet}
P.~Viana, astro-ph/0009492

\bibitem{Linde-h}
A.~D.~Linde, \plb{259}{1991}{38}

\bibitem{bcd00}
W.~Buchm\"uller, L.~Covi and D.~Del\'epine, \plb{491}{2000}{183}.

\bibitem{po77}
J.~Polonyi, Budapest preprint KFKI-93 (1977)

\bibitem{dss94} 
G.~Dvali, Q.~Shafi and R.~K.~Schaefer, \prl{73}{1994}{1886}

\bibitem{rsg95}
L.~Randall, M.~Soljacic and A.~Guth, \npb{472}{1996}{377}

\bibitem{grs01} 
G.~German, G.~Ross and S.~Sarkar, \npb{608}{2001}{423}

\bibitem{st97}
E.~D.~Stewart, \prl{391}{1997}{34}, \prd{56}{1997}{2019};
L.~Covi and D.~H.~Lyth, \prd{59}{1999}{063515};
L.~Covi, D.~H.~Lyth and L.~Roszkowski, \prd{60}{1999}{023509};
G.~German, G.~Ross and S.~Sarkar, \plb{469}{1999}{46}

\bibitem{lc00}
D.~H.~Lyth and L.~Covi, \prd{62}{2000}{103504};
L.~Covi and D.~H.~Lyth, MNRAS 326 (2001) 885.
    
\bibitem{cmbfast}
U.~Seljak and M.~Zaldarriaga, \apj{469}{1996}{437};\\
http://physics.nyu.edu/matiasz/CMBFAST/cmbfast.html

\bibitem{Max01}
A.~T.~Lee et al., astro-ph/0104459

\bibitem{Boom01}
C.~B.~Netterfield et al., astro-ph/0104460

\bibitem{Dasi01}
N.~W.~Halverson et al., astro-ph/0104489

\bibitem{Dasi01bis}
C.~Pryke et al., astro-ph/0104490

\bibitem{Max01bis}
R.~Stompor et al., astro-ph/0105062

\bibitem{Boom01bis}
P.~de~Bernardis et al., astro-ph/0105296

\bibitem{wtz01}
X.~Wang, M.~Tegmark and M.~Zaldarriaga, astro-ph/0105091

\bibitem{CBI}
S.~Padin et al., \apj{549}{2001}{L1} 

\bibitem{gauss}
J.~H.~P.~Wu et al., astro-ph/0104248;
S.~F.~Shandarin et al., astro-ph/0107136;
M.~G.~Santos et al., astro-ph/0107588 

\bibitem{adiab} 
D.~Langlois and A.~ Riazuelo, \prd{62}{2000}{043504};
R.~Trotta, A.~Riazuelo and R.~Durrer, astro-ph/0104017

\bibitem{BBN00}
S.~Burles, K.~M.~Nollett and M.~S.~Turner, \apj{552}{2001}{L1}

\bibitem{omegam}
M.~S.~Turner, astro-ph/0106035

\bibitem{hann00}
S.~Hannestad, S.~Hansen and F.~L.~Villante, \app{16}{2001}{137}

\bibitem{hann01}
S.~Hannestad et al., astro-ph/0103047

\bibitem{map}
http://map.gsfc.nasa.gov

\end{thebibliography}
\end{document}